# The GaN yellow-luminescence-related surface state and its interaction with air

Yury Turkulets,[1] Nitzan Shauloff,[2] Or Haim Chaulker,[1] Yoram Shapira,[3] Raz Jelinek,[2] and Ilan Shalish[1]*

[1]*School of Electrical Engineering, Ben-Gurion University, Beer Sheva 8410501, Israel.*
[2]*Department of Chemistry, Ben-Gurion University, Beer Sheva 8410501, Israel*
[3]*Department of Physical Electronics, Tel Aviv University, Tel Aviv 6997801, Israel*

Yellow luminescence (YL) is probably the longest and most studied defect-related luminescence band in GaN, yet its electronic structure or chemical identity remain unclear. Most of the theoretical work so far has attributed the feature to bulk defects, whereas spectroscopic studies have suggested a surface origin. Here, we apply deep level spectroscopy using sub-bandgap surface photovoltage that provides the energy distribution of the surface charge density. Comparison of surface charge spectra obtained under identical conditions before and after various surface treatments reveals the dynamics of the surface charge density. Further comparison with spectra of the entire state obtained using photoluminescence shows how the charge density stored in YL-related defects is eliminated upon a mild anneal in vacuum. This suggests that the YL-related defect involves a certain molecule adsorbed on the GaN surface, possibly in a complex with an intrinsic surface defect. The observed interaction with air strongly indicates that the YL-related deep level is a surface state.

## I. INTRODUCTION

While Si technology has been departing from Moore's law, GaN has been emerging as the next technological semiconductor after silicon and has already secured its superiority in the niche of power electronics. Power electronics has been coming into focus with the advent of electric vehicles.[1] One of the major concerns in transistor semiconductors is crystal defects, and much research is naturally devoted to find ways to deactivate them. Defects have an undesired effect on device performance because they trap charge. Trapped charge affects the operational stability of GaN-based devices. Probably the best example for such instability is observed in AlGaN/GaN high electron mobility transistors (HEMT). AlGaN surface states and GaN bulk states are known to be the main source of current collapse and $R_{DS-ON}$ issues during the transistor operation.[2,3] Perhaps the most common of the GaN defects is the one associated with yellow luminescence (YL).[4] Despite its limitations, photoluminescence (PL) has been the most common method in the study of semiconductor defects due to its ease of use and availability. An exceptionally wide sub-bandgap peak is often observed in PL spectra of GaN centered at ~2.2 eV. This wide peak is usually referred to as a "band" and has been dubbed the Yellow Band for the strong yellow emission often observable to the naked eye that governs the PL emission from GaN.

One of the intensive research debates regarding the YL has been the question whether it is a *surface state or a bulk state*.[5,6,7,8,9,10,11,12,13,14,15,16,17,18,19] The electronic device industry places particular value on the answer to this question, because surface states are accessible to surface treatments and hence may be more readily deactivated compared with their bulk counterparts.

Over 1400 studies have been published on the origin of YL over the course of several decades. Theoretical ab-initio studies have been traditionally limited to bulk. Several such studies suggested the source to be Ga vacancies.[5,6,7] YL was also observed in p-type Mg-doped GaN and was also attributed to N-vacancies. Van de Walle and Segev, presented an ab-initio model for surface state characterization and used it to study the surface of GaN.[9] Their results do not point to any YL-related state, and the model in general suggested similar energetic configuration for bulk and surface defects. In all, first principle calculations have not been able to define unambiguously whether the YL was a bulk or a surface defect. Experimental attempts to identify the exact location of the related defect have been numerous as well. In one of the earliest studies, Pankove and Hutchby studied PL in GaN that was ion implanted with 35 different elements, most of them induced a luminescence peak centered at ~2.15 eV.[10] While it seems likely that the damage of this implantation created additional Ga-vacancies in the bulk, the surface should have been damaged as well at the same time, thereby increasing the surface state density. Several studies suggested that a carbon impurity produces the YL.[11,12,13,14,15]

Regardless of their exact energy distribution, it is widely accepted that c-plane GaN has a high concentration of surface states.[20] Charge trapped in these states partially compensates the polarization charge. The result is upward band bending at both the Ga and the N faces of GaN, despite the opposite sign of their polar charges.[21,22] In GaN HEMTs, high density of surface states actually benefits the device as the source of the two-dimensional electron gas (2DEG).[23,24,25] Several spectroscopic studies have shown a clear relation of the YL band to surface states. Cathodoluminescence studies correlated YL to low angle grain boundaries which contain dislocations.[16] PL and surface photovoltage spectroscopy (SPS) identified the YL-related defect as a deep acceptor probably located at the surface.[17] A study of Pt/n-GaN Schottky barrier using internal photoemission spectroscopy showed photon-induced unpinning of the surface Fermi level with a substantial increase of the Schottky barrier height.[26] These results support the hypothesis of the YL surface origin. Based on PL, Hall effect, and deep level transient spectroscopy (DLTS) a transition from a shallow donor level to a deep level at $E_v + 0.87$ eV was suggested to underpin the YL emission.[18] While not specifically concerned with the question of surface vs. bulk origin, they did show that dry etch modified surface defects and increased the intensity of the yellow band, and suggested the formation of $O_N$ (oxygen occupying a nitrogen site) complexes at the surface. A PL study of triangular cross-section GaN nanowires (NWs) showed that a feature at 2.26 eV was associated with surface states.[19]

The main concern for device performance is charge trapping by surface states. A useful technique for characterizing charge trapped in defects is surface photovoltage spectroscopy (SPS).[27] Foussekis et al. used SPS to study the effect of UV illumination on the surface band bending of GaN under various ambient conditions.[28] They showed that UV illumination of air exposed GaN induced chemisorption of a certain substance on its surface, while illumination of the same sample in vacuum caused desorption of the adsorbates from the surface. Repeating the same experiment in dry nitrogen and oxygen, instead of air, they suggested that the adsorbate responsible for the reported behavior was oxygen. It was also shown that the surface band bending depended on the sample temperature, suggesting desorption of surface species at elevated temperatures in vacuum during UV illumination.[29] Thus, while quite a few studies pointed to a surface origin of the YL, no clear-cut experimental evidence has been put forward as of yet.

Here, we use a charge characterization technique based on SPS to study the charge trapped at the surface of GaN. Our model provides qualitative energy distribution of the surface charge density over the GaN sub-bandgap energy range. We apply a series of different treatments to explore the dynamics of the surface charge distribution (occupancy) of the YL-related state. We show how this state (observed by PL) may be fully depleted of its charge (observed by SPS) using a mild thermal treatment in vacuum and then fully recharged on exposure to air.

## II. MODEL

To obtain the spectrum of electrical charge from surface photovoltage spectra, one has to model the relation between them. This will eventually lead us to show that we have to take a photon-energy derivative of the surface photovoltage spectrum. The first attempts to use the photon-energy derivative of the photovoltage were made by Lagowski *et al.*[30] However, they have mistakenly assumed that the result of their model provided the dependence of the photoionization cross-section on the illumination photon energy.[27] Kronik *et al.* proposed a model suggesting that the energetic distribution of the surface charge can be obtained from the energy derivative of the square root of the photovoltage.[31] If the surface state distribution is fully occupied (charged) the distribution of surface charge and the distribution of the surface state are identical. However, as we will see here, surface treatments may change the charge distribution considerably, and therefore, to explore their effect, two methods are required. One, to observe the distribution of the state, and another, to observe the distribution of charge within this state. For the former, we use PL. For the latter, we use the SPS derivative. The model presented here uses a similar approach to obtain a *qualitative* energy distribution of the surface charge. In principle, the following model can provide a fully *quantitative* surface charge distribution. However, this would require the knowledge of several parameters, of which the precise experimental evaluation is relatively difficult. To support the conclusion that we propose here on the surface origin of the YL-related states, we believe a *qualitative* version of the model suffices, while a fully quantitative treatment may only add a marginal support if at all, in disproportion to the required effort.

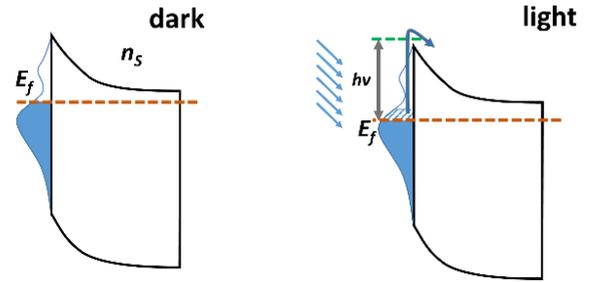

FIG 1 Cartoon showing a partially charged surface state and the typical change in its charge distribution upon illumination Comparing the distribution a surface state with that of its charge, as in this cartoon, could make a powerful tool for the study of surface state dynamics. In this paper, we propose a method to this end.

Without limiting the generality of the model, we will limit our discussion to n-type semiconductors. Abrupt termination of the semiconductor lattice, along with interaction of the surface with ambient substance may introduce allowed states within the forbidden gap. These states may either be vacant or occupied by charge carriers. The exact part of the surface states that is occupied (or charged) is defined by the Fermi-Dirac distribution. For simplicity, we will assume that all the states below the Fermi level energy are occupied by electrons, while those above it are vacant. Typically, the origin of the charge trapped in surface states is in electrons depleted from the bulk close to the semiconductor surface. This depletion results in an upward surface band bending in equilibrium. If, for example, one of these surface trapped electrons is excited using sufficiently energetic photon to be emitted over the surface barrier back into the bulk, the surface band bending will decrease (**Fig. 1**). It is difficult to evaluate the band bending without disturbing the equilibrium.[32,33] However, the change in the band bending caused by photo-excitation of charge over the surface barrier is readily evaluated by measuring the change in the contact potential difference, i.e. the surface photovoltage.

The density of bulk states, whether lattice imperfections or impurities, is generally lower by few orders of magnitude than the typical density of surface states. This is because the discontinuity of the lattice at the surface potentially provides a defect site per almost every surface atom. Moreover, not all bulk states affect the surface photovoltage, only those within the thin depletion layer. This further reduces their effect on the photovoltage as compared to that of surface states.

Sub-bandgap photovoltage can be caused by either of two possible processes: (1) excitation of charge from a surface state to a band, or (2) excitation of charge from a bulk state, into one of the bands within the surface depletion layer. For the case of the first process, the change in the surface band bending upon illumination, $\Delta V_{BB}(h\nu)$, caused by charge excitation over the surface barrier at a specific photon energy, $h\nu$, can be defined in terms of the Poisson equation:

$$V_{BBD} - \Delta V_{BB}(h\nu) = \frac{q}{2\varepsilon N_D}\left(N_{TD} - \Delta N_T(h\nu)\right)^2 \quad (1)$$

where $V_{BBD}$ is the surface band bending at equilibrium (or in the dark), $N_{TD}$ – the surface charge density at equilibrium, $\Delta N_T(h\nu)$ – the change in the surface charge, $N_D$ – carrier (donor) concentration, $q$ – electron charge, and $\varepsilon$ – dielectric constant. Taking the derivative of Eq (1) with respect to the photon energy, $h\nu$, yields:

$$\frac{d\Delta V_{BB}(h\nu)}{dh\nu} = \frac{q}{\varepsilon N_D}\left(N_{TD} - \Delta N_T(h\nu)\right)\frac{d\Delta N_T(h\nu)}{dh\nu} \quad (2)$$

Substitution of (1) into (2) yields the density of occupied states at a specific photon energy (or a specific surface energy with respect to the conduction band at the surface) $N_{OC}(h\nu)$.

$$N_{OC}(h\nu) = \frac{d\Delta N_T(h\nu)}{dh\nu} = \frac{c}{\sqrt{V_{BBD} - \Delta V_{BB}(h\nu)}}\frac{d\Delta V_{BB}(h\nu)}{dh\nu} \quad (3)$$

This equation requires 2 parameters which in principle may be obtained experimentally: $V_{BBD}$ and c. To obtain c, one has to measure the carrier concentration in the sample. However difficult, it should also be possible to obtain $V_{BBD}$ experimentally and several methods have been previously proposed to that end.[33] We could also leave them as unknowns and evaluate their effect on the expected result. The term $\sqrt{V_{BBD} - \Delta V_{BB}(h\nu)}$ depends on the surface band bending in the dark (equilibrium band bending), which is unknown. However, we can approximate its value by illumination with high intensity and measuring the resulting photovoltage (the so-called photosaturation technique).[34] The photon energy of the pumping beam should be high enough to excite the entire spectrum of surface states but lower than the band gap energy to prevent band-to-band excitation arising due to the Franz-Keldysh effect. Typically, a photon energy 200 meV below the band gap energy should suffice to prevent Franz-Keldysh excitation.[35] The resulting photovoltage may not equal the surface band bending in the dark, but should give a lower boundary of its value. If this result is significantly higher than the photovoltage values obtained by the illumination intensity used in our experiment, i.e. $\Delta V_{BB}(h\nu) \ll V_{BBD}$, we may assume the term $\sqrt{V_{BBD} - \Delta V_{BB}(h\nu)}$ to be constant, i.e. we assume a small perturbation regime. This also means that the photon flux used in our measurements does not excite the entire density of the surface charge but rather a small part of it. Therefore, to obtain the total density of the surface charge we should multiply our result by a constant factor, which represents the portion of the surface charge density that is excited by the photon flux used in our measurement. We now combine all these constants under C, which contains, among the elementary charge and material specific parameters from (1) and (2), the portion of the total charge excited by our illumination intensity. Assuming that this number remains constant throughout the entire spectrum we may define:

$$N_{OC}(h\nu) \approx C\frac{d\Delta V_{BB}(h\nu)}{dh\nu} \quad (4)$$

Using Eq. 4, we can get a qualitative energy spectrum of the surface charge, which is relative to the actual value, even if its y-axis is not calibrated. Nonetheless, this suffices for our needs in the present study. Even if C does change mildly with the photon energy, it should be safe to assume that its variations are the same among several spectra acquired with the same setup on the same sample. We can thus quantitatively compare *qualitative* spectra obtained from the same specific sample under the same conditions but after different surface treatments.

### III. MATERIALS AND METHODS

The GaN used in this study was a 2.2 um thick metal organic vapor phase epitaxy grown n-type Ga-face c-plane GaN on sapphire doped with both Si and Mg. All measurements were carried out inside a stainless-steel vacuum chamber attached to a Janis ST-500 cryostat equipped with a heatable stage. To reduce the effect of noise on the derivative of the photovoltage spectra, we carried out the experiments on this temperature stabilized sample stage,. The stage temperature was controlled using a Lake Shore 330 autotuning temperature controller and

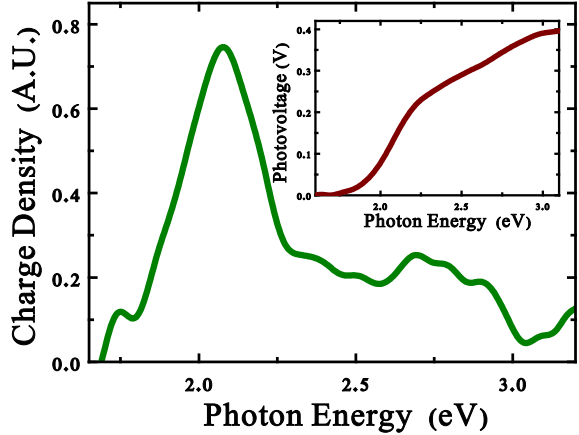

FIG 2 Surface charge density obtained from the photovoltage spectrum shown as inset. For clarity, absolute values of both plots are shown. It exemplifies the use of the model presented here to obtain a qualitative charge density spectrum. The model assumes that the photovoltage is caused by surface states. The validity of this assumption for the YL-related state is established in the following experiment.

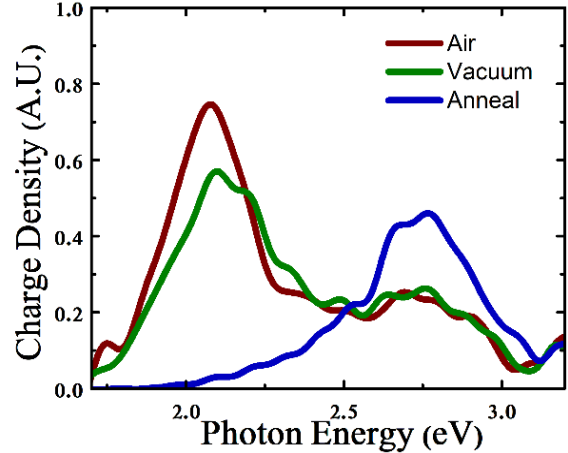

FIG 3 Surface charge density obtained from same GaN sample at room temperature: at air (red), in vacuum (green), and after heating up to 450K for 24 hours in vacuum (blue).

## IV. RESULTS AND DISCUSSION

**Fig. 2** shows a surface charge density profile along with the photovoltage spectrum from which it was calculated using the above method. A charge distribution peak is observed over part of the YL band energy range, centered at around 2.08 eV. The distribution of charge may be different from the distribution of the state, because the state is not always charged in its entirety. For this reason, the charge density spectrum peaks at a slightly different energy than the typical PL spectrum.

Three SPS spectra were acquired from the same GaN sample at room temperature before and after the following treatments: (1) The sample was first measured in air, and then (2) after evacuation of the measurement chamber to high vacuum (base pressure lower than $1 \times 10^{-5}$ Torr), (3) following a vacuum heat treatment at 450 K for 24 hours and cooling down back to room temperature. Surface charge distribution spectra were then calculated from these photovoltage spectra using the above model and are shown in **Fig. 3**.

The charge spectrum obtained from GaN in air has two distinct features: a sharp peak centered at 2.08 eV (YL) and another wide peak at 2.77 eV (blue luminescence - BL). The magnitude of the YL peak decreases when measured after the evacuation of the chamber, while the BL peak remains unchanged. Following the anneal at 450 K the YL peak is totally absent from the spectrum, while the BL increases its magnitude. These results suggest that the origin of the YL-related charge is probably in a certain surface adsorbate on the GaN surface, which gradually desorbed during the experiment. Evacuation of the sample environment seems to cause only limited desorption of molecules that are weakly attached to the surface, while anneal at 450 K desorbs all the remains of this adsorbate. Interestingly, this process was found to be fully reversible. Exposure of the sample to air restored the charge was set to 305K in all measurements. Vacuum annealing was carried out on the same stage. We used a mild annealing temperature of 450K in high vacuum to avoid any major chemical changes and stay as much as possible from lattice decomposition temperatures, while still being able to cause desorption.

Surface photovoltage was measured using a Kelvin probe (Besocke Delta Phi Gmbh). The sample was illuminated using a xenon light source monochromatized by a Newport MS257 spectrometer and filtered by order-sorting filters. At each wavelength step, the sample was illuminated for 5 min, at the end of which electrical measurements took place. The steps were equally spaced in energy (5 meV). At each point, the acquired value was obtained by averaging 100 consecutive measurements. The total time to acquire each spectrum was 33 hrs and it followed a relaxation period of 48 hrs in which the sample was kept in the dark in the Faraday cage chamber to make sure that the sample is at equilibrium before spectral acquisition commences. All the photovoltage spectra were obtained under a constant photon flux. The flux was maintained constant using an automatically-controlled slit at the light source. It is important to control the flux during the acquisition rather than normalizing the spectra after the acquisition, because the latter method assumes that the optical response is linear with the light intensity while it is rarely so.

PL was excited using a 1 mW He-Cd laser (Meles-Griot Ltd.) lasing at 325 nm and acquired using a Newport MS257 spectrometer with a Si CCD detector.

density spectrum to its original distribution, including the YL-related feature. This cycle was reproduced 3 times.

The shape of the BL, remains unchanged in the first two spectra and increases in magnitude after the mild anneal. This suggests that its origin is probably not in weakly adsorbed molecules. While it may be a bulk state, it may also be a surface state that is not affected by our treatments. In fact, a spectral feature in this exact range has been previously attributed to surface oxidation of GaN.[36] Oxidation is a chemical process, whose product may require much higher energy to be removed, much greater than could be provided by the mild anneal used here. While we cannot conclude about its origin, the BL appears to be an independent spectral feature located at energies higher than the YL. Therefore, its presence has no bearing on the conclusions we wish to draw regarding the origin of the YL.

The reason for the increase in the BL magnitude after the anneal appears to be technical. We compare different spectra under the assumption that the band bending in the dark does not change significantly with our surface treatments. However, a complete elimination of the YL charge, which by rough estimate makes up at least half of the total surface charge density, if not more, must introduce a significant change to the surface band bending in the dark. If this change is not properly factored in, it might introduce an error in the form of an increase of the peak magnitude.

The SPS results of this study suggest that the charge stored in the YL-related state is removed upon a mild anneal in vacuum and restored after exposure to air. This does not necessarily imply that the YL-related state desorbs entirely. This is because SPS can only sense the surface charge and thus can give the distribution of the charged states only. To sense all the surface states, whether or not occupied, we use PL. We, therefore, acquired PL spectra from same GaN sample under the exact same conditions of the SPS experiment: in air, in vacuum, and after anneal in vacuum at 450K for 24h followed by cooling down to RT. The PL spectra (**Fig. 4**) show a practically unchanged magnitude of the YL peak, suggesting that the defect associated with this spectral feature is not desorbed. This important observation suggests that only the *charge* associated with the YL-related defect comes from an air-constituent molecule. This molecule adsorbs onto the surface state in air and desorbs by the vacuum anneal.

Comparing the YL-related peaks obtained from PL and SPS, it is evident that the latter is narrower and appears slightly shifted to a lower photon energy than the former. Narrower energy distribution of the charge distribution (SPS) compared to the surface state distribution (PL) indicates that the related state is only partially occupied. Whereas the shift of the surface charge distribution to lower energy appears to be due to the presence of the surface band bending barrier during the photovoltage acquisition. In the PL measurement, the high power of the exciting laser beam practically flattens the bands at the surface while strongly evacuating the state of its charge. Therefore, the PL spectrum provides the actual energy distribution of the surface state.[25] However, in SPS, the small perturbation regime leaves in place a significant surface band bending. Electrons surpass this band-bending barrier mostly by internal photoemission. However, the built-in electric field within the band-bending region gives rise to a certain level of field emission through the barrier, i.e. *field-assisted absorption*, also known as the Franz-Keldysh effect.[35] This way some electrons traverse the barrier at less than its full height, thereby red-shifting the resulting charge spectrum (to a degree that is dependent on the field strength).

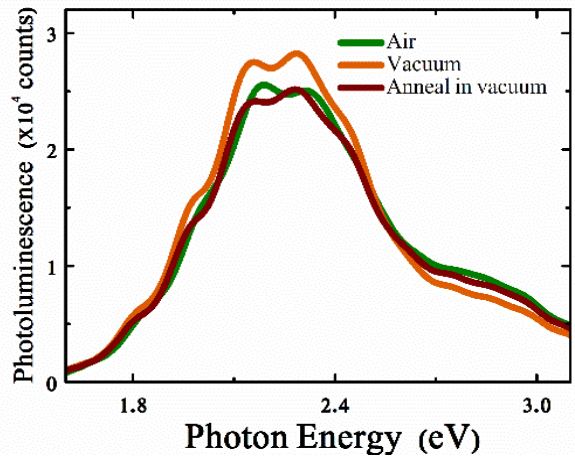

FIG 4 Photoluminescence spectra of the GaN at air, in vacuum and after anneal in vacuum at 450K for 24h. No significant change is observed after the various treatments, suggesting that the YL-related state does not desorbs, only the charge associated with it.

The present results may answer the question regarding the location of the YL-related state. In general, YL can be associated with two different locations within the layer: bulk, or surface. Removal of the YL-related feature from the charge density spectrum after a mild anneal and its restoration on exposure to air present a clear scenario of interaction with the environment. A deep level situated in the bulk cannot interact with the environment without significant outdiffusion. Since outdiffusion is not possible at the low aneal temperature used here, the only possible configuration that supports interaction with the environment requires that the deep level will be situated at the border with the environment, i.e., at the very surface. This study thus provides the most direct experimental evidence so far that the YL-related state is a *surface state*.

Since this study was carried out on one of the polar faces of GaN, it raises a question: Does the polar charge, which is practically only at the very surface, has a role in this adsorption? In the more typical non-polar semiconductor surface, the surface trapped charge originates in a mobile charge from the bulk that is surface-trapped during the formation of the surface depletion layer. GaN is slightly different from the common semiconductor, because it is polar. This study was carried out on a c-plane of a GaN epilayer grown on sapphire, having only its Ga-face exposed to air. The Ga-face has a negative polar charge, which is able to induce a surface depletion layer due to Coulomb repulsion of free electrons from the surface vicinity deep into the bulk. At the same time, this large polar charge prevents free electrons from reaching the surface and being trapped in surface states. The other GaN surface (the N-face), interfacing the sapphire substrate, has a positive polar charge, which attracts electrons to that surface. Depending on the layer thickness and the doping density, this uneven distribution of free electrons can compensate the polar charge on both surfaces getting the layer into equilibrium. Nonetheless, since the two GaN surfaces have opposite polar charges, the electric field emanating from these polar charges can only exist within the layer. Therefore, electrostatic attraction of charged molecules from the environment to surface polar charge is not likely. The driving force behind the observed surface adsorption is therefore likely to be of a different kind, e.g. minimization of the surface energy by reaction of one of the air constituents with dangling bonds.

In fact, semiconductor free surfaces have long been known to interact with ambient gasses. Chakrapani has shown that when ambient $O_2$ and $H_2O$ are adsorbed on surfaces of various semiconductors, they may act as a deep acceptor leading to the formation of a surface depletion region.[37] GaN makes no exception, and several adsorption studies have also been carried out on GaN surfaces. Based on first principle calculations, Rapcewicz et al. suggested that the large electronegativity of N in GaN provides adsorption sites at the GaN surface.[38] In their paper, they also showed that adsorption of hydrogen on the GaN surface, regardless of the surface polarity, stabilizes the surface morphology. On the other hand, Bermudez has shown experimental evidence that oxygen can be chemisorbed on the GaN surface.[39] Multiple other studies have also reported adsorption of various materials on GaN.[40,41,42,43,44,45] However, as of yet, neither these studies nor our present results can unambiguously associate a specific air molecule with the YL.

## V. CONCLUSION

Our results reveal the power of a combined method for observing deep level spectra (by PL) and the distribution of their charge (by SPS). Exploring deep level charge dynamics in response to a surface treatment, we show that the YL-related state may be depleted of its charge by a mild anneal in vacuum and then fully restored to its initial distribution by exposure to air. The conclusions that emerge seem to be compatible only with the premise that the GaN YL-related deep level is a surface state that interacts with airborne molecules. These results are important news for the GaN device community because a surface state is accessible to surface treatments and may be removed or passivated. This surface charge may be responsible for some of the instabilities observed in GaN devices and our findings can pave the way to its prevention or deactivation.

**Acknowledgment**

Financial support from the Office of Naval Research Global through a NICOP Research Grant (No. N62909-18-1-2152) is gratefully acknowledged. DISTRIBUTION A. Approved for public release, distribution unlimited. **DCN# 43-10073-22**